\documentstyle[12pt,aps,prc]{revtex}
\newcommand{\be}{\begin{equation}}
\newcommand{\ee}{\end{equation}}
\newcommand{\bel}[1]{\be\label{#1}}
\newcommand{\re}[1]{Eq.~(\ref{#1})}

\newcommand{\qq}{\mbox{$q \overline{q}$}\,\,}
\newcommand{\rhob}{\overline{\rho}}
\newcommand{\goo}{\,\raisebox{-.5ex}{$\stackrel{>}{\scriptstyle\sim}$}\,}     
\newcommand{\loo}{\,\raisebox{-.5ex}{$\stackrel{<}{\scriptstyle\sim}$}\,}
\newcommand{\bm}[1]{\mbox{\boldmath${#1}$\unboldmath}}
\newcommand{\sump}{\sum_{\bm{p}}}
\newcommand{\sumpl}{\sum_{\bm{p},\lambda}}
\newcommand{\ocn}{n_{\bm{p}}}
\newcommand{\ocnb}{\overline{n}_{\bm{p}}}

\begin{document}
\draft
\renewcommand{\thefootnote}{\arabic{footnote}}
\begin{center}
{\Large\bf Metastable quark--antiquark droplets\\
within the Nambu--Jona-Lasinio model}\\[5mm]
{\bf I.N.~Mishustin$^{\,\ast,\,\dagger}$, L.M.~Satarov$^{\,\ast,\,\ddagger}$, 
H.~St\"ocker$^{\,\ddagger}$ and W.~Greiner$^{\,\ddagger}$}
\end{center}
\begin{tabbing}
\hspace*{2.5cm}\=${}^\ast$\,\={\it The Kurchatov~Institute,
123182~Moscow,~\mbox{Russia}}\\
\>${}^\dagger$\>{\it The Niels~Bohr~Institute,
DK--2100~Copenhagen {\O},~\mbox{Denmark}}\\
\>${}^\ddagger$\>{\it Institut~f\"{u}r~Theoretische~Physik,
J.W.~Goethe~Universit\"{a}t,}\\
\>\>{\it D--60054~Frankfurt~am~Main,~\mbox{Germany}}
\end{tabbing}

\begin{abstract}
Chemically non--equilibrated quark--antiquark matter is studied within the
Nambu--Jona-Lasinio model.  The equations of state of non--strange
(\mbox{$q=u,\,d$}) and strange (\mbox{$q=s$}) 
$q\overline{q}$ systems are calculated in the
mean--field approximation. The existence of metastable bound states 
with zero pressure is predicted at finite densities and 
temperatures $T\loo$ 50 MeV. 
It is shown that the minimum energy per particle occurs
for symmetric systems, with equal densities of quarks and antiquarks.
At $T=0$ these metastable states have quark number densities of about 
\mbox{0.5 fm$^{-3}$} for \mbox{$q=u,\,d$} and 
of \mbox{1 fm$^{-3}$} for \mbox{$q=s$}. 
A first order chiral phase transition is found at finite densities and
temperatures.  The critical temperature for this phase transition
is approximately 75~MeV (90 MeV) for the non--strange (strange) baryon--free
quark--antiquark matter. For realistic choices of parameters, 
the model does not predict a phase transition in chemically 
equilibrated systems. 
Possible decay channels of the metastable \qq droplets 
and their signatures in relativistic heavy--ion collisions are discussed.  
\end{abstract}

\pacs{PACS numbers: 12.38.Mh, 11.30.Qc, 12.39.Fe, 25.75.-q}
%
%


\section{Introduction}

Studying the equation of state (EOS) of strongly interacting
matter is necessary for understanding 
the evolution of the early universe, properties of neutron stars 
and dynamics of heavy--ion collisions. 
First of all, one should know the degrees of freedom
which are most relevant at a given energy density.  
When studying the EOS it is also important   
to identify the regions of possible phase transitions and
to find all stable and metastable states which correspond to extrema 
of a thermodynamic potential.
Presumably, lattice QCD calculations can be used to
study these questions from ``first principles''.
At present, however, this approach gives reliable 
results only for a baryon--free matter in 
thermodynamic equilibrium.

Therefore, one should use effective theories when 
dealing with a wider class of multiparticle systems.  
The Nambu---Jona-Lasinio (NJL) \mbox{model~\cite{Nam61,Lar61}} 
is  one of such effective theories
which proved to be rather successful in describing
the ground states of light hadrons in vacuum. 
The chiral symmetry of the strong interactions, most 
important for the low energy hadron physics, is explicitly
implemented in this model. Although the model is non--renormalizable
and does not contain gluons
\footnote{
Lattice calculations show that at temperatures 
$T\loo 150$\,MeV the gluons acquire a large effective mass due to the color 
screening effects. In the present work, dealing mostly with such moderate 
temperatures, we assume that gluonic degrees of freedom are 
suppressed.
}, it seems to be a reasonable approximation to QCD  
at scales intermediate between asymptotic freedom and confinement. 
   
In this paper the NJL model is applied to investigate the properties
of chemically non--equilibrated quark--antiquark (\mbox{$q\overline{q}$}) 
matter. The study of possible bound states and 
phase transitions is one of the main goals of the present work.
The search for new states of strongly interacting systems, essentially
different from normal nuclear matter, has already quite
a long history. Using the linear $\sigma$--model, Lee and Wick
predicted~\cite{Lee79} a metastable superdense state of nucleonic matter
at $T=0$. This problem was further studied in Ref.~\cite{Bog83}
by including the repulsive vector interaction. 
The liquid--gas as well as the chiral phase transitions 
of hadronic matter have been 
investigated within the framework of the relativistic mean--field model in 
Refs.~\cite{Wal85,The83,Mul95}. Since the chiral symmetry is not respected
by this model, its predictions become questionable
at high baryon densities and temperatures. 
Attempts to generalize the hadronic models by 
implementing the main symmetries of QCD have been made in 
Refs.~\cite{Ell92,Mis93,Pap97}.
A chiral phase transition was found in nuclear matter 
only at very high baryon densities and temperatures when hadronic 
degrees of freedom are inappropriate. 

The existence of some exotic multiquark and multihadron states
has been discussed by various authors. For example, 
``stranglets'' (bound states of strange and light quarks) have been proposed 
in Ref.~\cite{Far84}. The possibility of their formation in relativistic 
heavy--ion collisions has been considered in Ref.~\cite{Gre87}. 
Arguments that  multipion systems
may be bound have been given in Ref.~\cite{Mi93a}.
The phase transition of quark matter into a color superconducting
state has been predicted recently~\cite{Rag98,Shu98} within the 
QCD--motivated models. The random matrix model has been used to study 
the phase diagram of quark matter in Ref.~\cite{Jak98}.
Possible signatures of QCD phase transitions have been discussed
in Ref.~\cite{Ste98}.

The possibility of multiquark bound states  and chiral phase transitions 
was studied by many authors (see e.g.~\cite{Asa89,Kli90a,Cug96,Kli97}) 
within the NJL model. Attempts to apply the same model to  
nucleonic matter were made in Refs.~\cite{Koc87,Bub96,Mis96}. 
Most of these studies have been done within the 
mean--field approximation. An attempt to go beyond this approximation, 
by including the mesonic degrees of freedom, has been made 
in Ref.~\cite{Zhu94}. The results of these works, obtained
with different sets of model parameters, often contradict to each other. 
A common shortcoming appearing in many papers 
is the omission~\cite{Cug96,Kli97} or underestimation~\cite{Asa89}   
of the repulsive vector interaction\footnote{
The vector interaction was not considered in the original
version of the model~\cite{Nam61}.
}. 
It was shown in Ref.~\cite{Kli90a} that
this interaction becomes rather important at large baryon densities.
According to Ref.~\cite{Kli90a}, at realistic values of the vector 
coupling constant the NJL model does not predict 
a first order phase transitions in quark matter.  

All these results have been obtained by assuming that the 
conditions of statistical equilibrium are perfectly fulfilled in the matter.
This implies two kinds of equilibration. 
The first, kinetic or thermal equilibrium, means that the occupation numbers
of quarks and antiquarks coincide with the equilibrium distribution functions
characterized by some temperature and chemical potentials.
The second, chemical equilibrium, requires that the certain relations
between chemical potentials of different particles must be 
fulfilled. These relations determine the equilibrium abundances of 
various species which can be reached at large times due to inelastic 
multiparticle interactions. First attempts to study non--equilibrated 
\qq systems were made in Refs.~\cite{Zha92,Aba95,Mis97} by using
transport equations derived from the NJL model.

Off--equilibrium effects should inevitably accompany the formation 
and evolution of quark matter in relativistic heavy--ion collisions. 
There are many theoretical as well as experimental arguments 
in favour of large deviations from the chemical equilibrium, even 
in central collision of heaviest nuclei at c.m. bombarding energies 
$\sqrt{s}\goo 10$\, GeV per nucleon (i.e. for SPS energies and 
higher). This is a consequence of a short time available
for the interaction of primary nuclei and for subsequent 
equilibration of particles at such high energies. The degree 
of the chemical equilibration has been investigated~\cite{Gei93,Sri97} 
on the basis of the parton cascade model. Strong deviations from chemical 
equilibrium were predicted for light quarks at intermediate 
stages of a heavy--ion collision at RHIC and LHC energies. 
Large nonequilibrium effects 
at SPS energies were found within the UrQMD model~\cite{Bra98}.
As follows from the thermal model analysis of experimental data~\cite{Bra98b},
the hadronic matter is out of chemical equilibrium at late stages 
of relativistic nuclear collisions. 
The conclusion about a significant ``overpopulation'' of 
light quarks in 160 AGeV Pb+Pb collisions has been made in Ref.~\cite{Raf98}. 
  
Motivated by these findings, below we study the properties
of chemically non--equilibrated \qq matter 
which is however in the thermal equilibrium.
Formally, we introduce two chemical potentials 
for quarks and antiquarks with a given flavour and assume 
that their values do not  
satisfy in general  the conditions of chemical equilibrium. It will be
shown below that the EOS of chemically non--equilibrated \qq matter 
is non--trivial and its phase structure is much richer 
as compared with the equilibrated matter.                  
 
In the next section we briefly formulate the SU(3) version of the NJL
model used in the present paper. In Sec.~3 this model is applied to
calculate the EOS of the non--strange and strange \qq matter
at arbitrary densities of quark and antiquarks. 
The numerical results are presented
in Sec.~4. Sec.~5 is reserved for discussion and summary.   

\section{Formulation of the model}

We proceed from the SU(3) version of the NJL model 
suggested in Ref.~\cite{Vog91}. Its interaction part is given by
the local coupling between the quark color currents.
After the Fierz transformation the color singlet part of the  
Lagrangian may be written as:
\begin{eqnarray}
{\cal L}=\overline{\psi}\,(i\,\hat{\partial}-\hat{m}_0)\,\psi
&+&G_S\sum_{j=0}^{8}\left[
\left(\overline{\psi}\,\,\frac{\displaystyle\lambda_j}
{\displaystyle 2}\,\psi\right)^2+
\left(\overline{\psi}\,\frac{\displaystyle i\gamma_5\lambda_j}
{\displaystyle 2}\,\psi\right)^2\right]\nonumber\\
&-&G_V\sum_{j=0}^{8}\left[
\left(\overline{\psi}\,\gamma_\mu \frac{\displaystyle\lambda_j}
{\displaystyle 2}\,\psi\right)^2+
\left(\overline{\psi}\,\gamma_\mu \frac{\displaystyle\gamma_5\lambda_j}        
{\displaystyle 2}\,\psi\right)^2\right],\label{lagr}
\end{eqnarray}
where~~$\lambda_1,\ldots,\lambda_8$~~are the SU(3)
Gell-Mann matrices in flavour space,~~$\lambda_0\equiv\sqrt{2/3}\,\bm{I}$
~and \mbox{$\hat{m}_0={\rm diag}(m_{0u},\,m_{0d},\,m_{0s})$} is the 
matrix of bare (current) quark masses. 
At $\hat{m}_0=0$ this Lagrangian is invariant with respect to the
\mbox{${\rm SU(3)_{\,L}}\otimes\,{\rm SU(3)_{\,R}}$} chiral transformation.
The term with $\hat{m}_0$ leads
to explicit breaking of chiral symmetry which is supposed to be small.
The relation 
\bel{coup}
G_S=2\,G_V
\ee
between the scalar ($G_S$) and vector ($G_V$) 
coupling constants follows from the QCD motivated initial 
Lagrangian~\cite{Vog91}.

In fact, the above model can be formulated in such a way that
$G_V/G_S$ is considered as an adjustable parameter, which may be
chosen by fitting the observed masses of the vector mesons. Different 
calculations use values of $G_V/G_S$ in the range 
0.5--1~\cite{Vog91,Lut92}. We would like to stress here 
that the choice $G_V=0$, frequently made in the literature,
is unrealistic for baryon--rich
\qq systems. In this case, the omission of  
the repulsive vector interaction results in an underestimation of 
pressure leading to incorrect results concerning the possibility of
phase transitions and the existence of (meta)stable states
(see also Ref.~\cite{Bub96}). On the other hand, 
as will be seen below, thermodynamic functions of baryon--free systems 
are insensitive to $G_V$. 

In the mean--field (Hartree) approximation, used in the present paper,
only the scalar and vector
terms of~\re{lagr} survive and the Lagrangian is diagonal in 
the flavour space:
\bel{mfal}
{\cal L}_{\rm mfa}=\sum_{f=u,\,d,\,s}{\cal L}_f\,,
\ee
where
\bel{lagrf}
{\cal L}_f=\overline{\psi}_f(i\hat{D}-m_f)\psi_f
-\frac{\displaystyle G_S}{\displaystyle 2}<\overline{\psi}_f\psi_f>^2\\
+\frac{\displaystyle G_V}{\displaystyle2}
<\overline{\psi}_f\gamma_\mu\psi_f>^2\,.
\ee
Here $\psi_f$ denote spinors for $u, d, s$ quarks,
$D_\mu=\partial_\mu+i\,G_V<\overline{\psi}_f\gamma_\mu\psi_f>$, angular        
brackets correspond to quantum--statistical
averaging and the constituent quark mass $m_f$ is determined by the gap
equations
\bel{gape}
m_f=m_{0f}-G_S<\overline{\psi}_f\psi_f>\,.
\ee
Due to the absence of a flavour--mixing interaction,  
all extensive thermodynamical functions (energy density, pressure etc.) 
are additive in quark flavour.
Therefore, in this approximation light ($f=u,\,d$) and 
strange ($f=s$) quark systems can be studied separately. 
In the following we consider isospin--symmetric
\qq systems, assuming equal numbers of $u$ and $d$ quarks 
(antiquarks), and disregard the difference between $m_{0u}$
and $m_{0d}$\,.    

The model parameters $m_{0f}, G_S, \Lambda$ (the ultraviolet cutoff 
in 3--dimensional momentum space) can be fixed by reproducing
the empirical values~\cite{Rei85} of quark condensates
\mbox{$<\overline{u}u>$}, \mbox{$<\overline{s}s>$} and the 
observed values of $\pi$ and $K$ decay constants $f_\pi, f_K$
\footnote{These decay constants are related to the bare quark masses
and the condensate densities by the GOR relations~\cite{Gel68}.
}
in the vacuum.
We choose the following values of the model parameters~\cite{Cas95}
\bel{modp}
m_{0u}=m_{0d}=7\,{\rm MeV},\, m_{0s}=132\,{\rm MeV},\, G_S=24.5\,
{\rm GeV^{-2}},\,
 \Lambda=0.59\,{\rm GeV}\,,
\ee
The calculation shows (see Sec.~3) that these parameters correspond to
the values 
\begin{eqnarray}
f_\pi=93\,{\rm MeV},\hspace{2em}&&<\overline{u}u>=(-230\,{\rm MeV})^3, 
\hspace{2em}m_u^{\rm vac}=300\,{\rm MeV}\,,\label{empv1}\\   
f_K=90\,{\rm MeV},\hspace{2em}&&<\overline{s}s>=(-250\,{\rm MeV})^3, 
\hspace{2em}m_s^{\rm vac}=520\,{\rm MeV}\,,\label{empv}
\end{eqnarray}
where $m_f^{\rm vac}$ is the constituent quark mass in the vacuum. 

\section{Equation of state of quark--antiquark matter}

In the mean field approximation the Lagrangian is
quadratic in the quark fields and the calculation of thermodynamic functions
is straightforward. Below the method of
Ref.~\cite{Asa89} is generalized for chemically non-equilibrated
systems. The Coulomb and surface effects are disregarded and 
the characteristics of \qq matter are assumed to be spatially 
homogeneous and time-independent. In this article we consider separately 
the systems with light ($f=u,d$) and strange ($f=s$) quarks. Unless 
stated otherwise, the flavour index $f$ will be omitted. 
The main feature of a chemically non--equilibrated system 
is the existence of two chemical potentials, i.e. for quarks ($\mu$)
and antiquarks ($\overline{\mu}$) which are in general not strictly 
related to each other. If 
the equilibrium with respect to creation and annihilation of
\qq pairs were perfectly fulfilled, the above
chemical potentials would satisfy the condition of chemical equilibrium,
\bel{cemc}
\overline{\mu}=-\mu\,.
\ee   
At given temperature $T$ and quark density $\rho$~\re{cemc}
fixes the density of antiquarks $\rhob$ (see Fig.~8). 
Below we study a general case considering quark and antiquark densities 
and accordingly, their chemical potentials, as independent quantities.   

The generalized expression for the partition function $Z$ of the thermally
equilibrated \qq system may be written as
\bel{stat}
Z={\rm Sp}\,e^{-\widetilde{H}/T}=e^{-\Omega/T}\,,
\ee
where
\bel{htil}
\widetilde{H}=H-\mu\,N-\overline{\mu}\,\overline{N}\,.
\ee 
Here $H$ is the Hamiltonian and  $N$ ($\overline{N}$) is the quark
(antiquark) number operator.
Up to an arbitrary additive constant the thermodynamic potential 
$\Omega$ is equal to $-PV$ where $P$ and $V$ are respectively 
the pressure and the volume of the \qq matter.
Below the quantum--statistical averaging of any operator $A$ 
is made in accordance with the expression:
\bel{aver}
<A>\equiv {\rm Sp}\,(A\,e^{-\tilde{H}/T})/Z\,.
\ee   

From the Lagrangian,~\re{lagrf}, one obtains the equation of
motion for the quark field $\psi$
\bel{dire}
(i\,\gamma^\mu\partial_\mu-m-G_V\rho_V\gamma^0)\,\psi=0,
\ee
where $m$ is the constituent quark mass and 
$\rho_V=<\overline{\psi}\,\gamma^0\,\psi>$ is the vector density.
It can be expressed through the densities \mbox{$\rho=<N>/V$} and 
\mbox{$\rhob=<\overline{N}>/V$} as 
\footnote{The baryonic density $\rho_B$ of the \qq system equals 
$1/3\rho_V$\,.}:
\bel{vecd}
\rho_V=\rho-\rhob\,.
\ee

After the plane wave decomposition ~\cite{Bjo64} of $\psi, \overline{\psi}$\,,
the quantum states of quarks and antiquarks can be specified by  
momenta $\bm{p}$ and discrete quantum numbers $\lambda$
(helicity and color). Let  
$a_{\bm{p},\lambda}$\,(\,$b_{\bm{p},\lambda}$) and
$a^+_{\bm{p},\lambda}$\,(\,$b^+_{\bm{p},\lambda}$) be the destruction
and creation operators of a quark (an antiquark) in the state 
$\bm{p},\lambda$\,. Then the operators $H,\,N$ and $\overline{N}$
can be expressed through the linear 
combinations of $a^+_{\bm{p},\lambda}\,a_{\bm{p},\lambda}$
and $b^+_{\bm{p},\lambda}\,b_{\bm{p},\lambda}$\,.
For example
\bel{qden1}
\frac{\displaystyle N}{V}=\sumpl a^+_{\bm{p},\lambda}\,a_{\bm{p},\lambda},
\hspace*{5mm}
\frac{\displaystyle\overline{N}}{V}=
\sumpl b^+_{\bm{p},\lambda}\,b_{\bm{p},\lambda}
\ee
In the quasiclassical approximation the sum over momenta 
may be replaced by the \mbox{3--dimensional} integral over momentum space:             
\bel{msum}
\sump=\frac{\displaystyle 1}{\displaystyle (2\pi)^3}\int d^{\,3}p\,.
\ee  

By using Eqs.~(\ref{lagrf})--(\ref{gape}) one can find
the Hamiltonian density of the \qq matter in the mean--field
approximation. The resulting expression
for the operator $\tilde{H}$ reads as 
\begin{eqnarray}
\frac{\displaystyle\tilde{H}}{\displaystyle V}&=&\sumpl
(E_{\bm{p}}-\mu_R)\,a^+_{\bm{p},\lambda}\,a_{\bm{p},\lambda}
+\sum_{\bm{p},\lambda}(E_{\bm{p}}-\overline{\mu}_R)\,
b^+_{\bm{p},\lambda}\,b_{\bm{p},\lambda}\nonumber\\
&&-\sumpl E_{\bm{p}}+\frac{\displaystyle (m-m_0)^2}{\displaystyle 2\,G_S}
-\frac{\displaystyle G_V\rho_V^{\,2}}{\displaystyle 2}\,,\label{hamd}
\end{eqnarray}
where $E_{\bm{p}}=\sqrt{m^2+\bm{p}^2}$ and $\mu_R,\,\overline{\mu}_R$
denote the ``reduced'' chemical potentials:
\begin{eqnarray}
\mu_R&=&\mu-G_V\rho_V\,,\label{rce1}\\
\overline{\mu}_R&=&\overline{\mu}+G_V\rho_V\,.\label{rce2}
\end{eqnarray}

In~\re{hamd} the gap equation
\bel{gap1}
m=m_0-G_S\rho_S\,,
\ee	
has been used. Here $\rho_S\equiv<\overline{\psi}\psi>$
is the scalar density of the \qq system. 
It can be expressed as
\bel{scad}
\rho_S=<\sumpl\frac{\displaystyle m}{\displaystyle E_{\bm{p}}}
\left(a^+_{\bm{p},\lambda}\,a_{\bm{p},\lambda}+
b^+_{\bm{p},\lambda}\,b_{\bm{p},\lambda}
-1\right)>\,.
\ee
Eqs.~(\ref{hamd}) and~(\ref{scad}) contain divergent terms. 
They originate from the negative energy levels 
of the Dirac sea~\cite{Vog91,Mis96}.
In the present paper  these terms are regularized 
by introducing the three--dimensional momentum cutoff  
\mbox{$\theta\,(\Lambda -|\bm{p}|)$} where 
$\theta\,(x)\equiv\frac{1}{2}\,(1+{\rm sgn}\,x)$. The cutoff momentum 
$\Lambda$ is considered as an adjustable model parameter.

The quantum--statistical averaging  of the operator 
$a^+_{\bm{p},\lambda}\,a_{\bm{p},\lambda}$
gives the occupation number of quarks in the state 
$\bm{p},\lambda$. Direct calculation shows that it coincides
with the Fermi--Dirac distribution for an ideal gas with the chemical
potential $\mu_R$:      
\bel{ocnq}
<a^+_{\bm{p},\lambda}\,a_{\bm{p},\lambda}>\equiv\ocn =
\left[\,\exp{\left(\frac{E_{\bm{p}}-\mu_R}{T}\right)}+1\,\right]^{-1}\,.
\ee
The analogous expression for the antiquark occupation number, 
$\ocnb=<b^+_{\bm{p},\lambda}\,b_{\bm{p},\lambda}>$,
is obtained by the replacement $\mu_R\to\overline{\mu}_R$\,.
At given densities of quarks and antiquarks, 
the chemical potentials $\mu_R$ and $\overline{\mu}_R$ 
are determined from the normalization conditions
\bel{norc}
\rho=\nu\sump\ocn,\,\,\,\,\rhob=\nu\sump\ocnb\,,
\ee
where $\nu$ is the spin--color degeneracy factor.
For a \qq system with single flavour 
\bel{degf}
\nu=2 N_c=6\,.
\ee

After introducing the ultraviolet cutoff,~\re{scad} takes the form
\bel{scad1}
\rho_S=\nu\sump\frac{m}{E_{\bm{p}}}\left[\ocn+\ocnb-
\theta\,(\Lambda-p)\right]\,.
\ee 
The explicit form of the gap equation is given by Eqs.~(\ref{gap1}),
(\ref{ocnq}), (\ref{scad1}). 
The physical vacuum ($\rho=\rhob=\rho_V=0$) corresponds 
to the limit $\ocn=\ocnb=0$. 
As discussed above, the pressure can be represented as
\bel{pres}
P=\frac{T}{V}\ln{Z}+P_0\,,
\ee
where the additive constant $P_0$ is introduced to ensure that the vacuum
pressure is zero, \mbox{$P_{\rm vac}=0$}. Direct calculation gives the result
\bel{pres1}
P=P_K+\frac{\displaystyle G_V\rho_V^{\,2}}{\displaystyle 2}-B(m)\,.
\ee
Here the first (``kinetic'') term in the r.h.s. coincides 
with the total pressure of ideal gases of quarks and antiquarks 
having constituent mass $m$ and chemical potentials 
$\mu_R$ and $\overline{\mu}_R$\,:
\begin{eqnarray}
P_K&=&\nu\,T\sump\left\{\,\ln{\left[1+
\exp{\left(\frac{\mu_R-E_{\bm{p}}}{T}\right)}\right]}+\ln{\left[1+
\exp{\left(\frac{\overline{\mu}_R-E_{\bm{p}}}
{T}\right)}\right]}\right\}\nonumber\\
&=&\nu\sump\frac{\bm{p}^2}{3E_{\bm{p}}}\,\left(\ocn+\ocnb\right)\,.            
\label{presk}
\end{eqnarray}

The third term in~\re{pres1} gives the ``bag'' part of the pressure.
The explicit expression for $B(m)$ can be written as
\bel{bagc}
B(m)=\Phi (m_{\rm vac})-\Phi (m)\,.
\ee
Here $m_{\rm vac}$ is the constituent quark mass in the vacuum and
\bel{phim}
\Phi (m)=\nu\sump E_{\bm{p}}\,\theta\,(\Lambda -p)-
\frac{\displaystyle (m-m_0)^2}{\displaystyle 2\,G_S}\,.
\ee
From Eqs. (\ref{pres1})--(\ref{bagc}) one may see that
$P_{\rm vac}=-B\,(m_{\rm vac})=0$, i.e. the condition of vanishing vacuum
pressure is automatically satisfied.
The expression for $\Phi (m)$ can be calculated analytically,
\bel{phim1}
\Phi (m)=\frac{\displaystyle\nu\,\Lambda^4}{8\,\pi^2}\,
\Psi\left(\frac{\displaystyle m}{\displaystyle\Lambda}\right)-
\frac{\displaystyle (m-m_0)^2}{\displaystyle 2\,G_S}\,,
\ee
where
\bel{psid}
\Psi(x)\equiv 4\int\limits_0^1 dt\,t^{\,2}\,\sqrt{t^{\,2}+x^2}
=\left(1+\frac{x^2}{2}\right)\sqrt{1+x^2}
-\frac{x^4}{2}\ln{\frac{1+\sqrt{1+x^2}}{x}}\,. 
\ee

As required by the thermodynamic consistency, the gap equation (\ref{gap1})
should follow from the minimization of the thermodynamic potential $\Omega$
with respect to $m$,
\bel{gap2}
\left(\partial_m P\right)_{T,\,\mu,\,\overline{\mu}}=
\frac{\displaystyle m_0-m}{\displaystyle G_S}
-\rho_S=0\,.
\ee 
Therefore, the quark mass is not an independent variable and                    
the pressure can be considered as a function of $T,\,\mu$ and
$\overline{\mu}$ only. The gap equation in vacuum is
\footnote{The vacuum will be stable with respect to
mass fluctuations if $\Phi{\,''} (m_{\rm vac})<0$\,.}
\bel{gapv}
\Phi{\,'}\,(m_{\rm vac})=0\,.
\ee
This equation was solved numerically with the values of model 
parameters from~\re{modp}. The calculations were performed separately 
for light ($f=u,d$) and strange ($f=s$) quark flavours. In this way one
obtains the vacuum masses and scalar densities given 
in \mbox{Eqs.~(\ref{empv1})--(\ref{empv})}. It is convenient to define
the bag constants $B_{0f}$ for each flavour separately,  
as the difference between the 
pressures of physical ($m_f=m_f^{\rm vac}$) and perturbative ($m_f=m_{f0}$)
vacua, i.e. $B_{0f}=B(m_{0f})$\,. 
These quantities are analogous to the bag constants introduced in
the MIT bag model. For the present choice 
of parameters, the bag constants are   
\bel{bagc1}
B_{0u}=B_{0d}\simeq 34.5\,{\rm MeV/fm^3},\,\,B_{0s}\simeq 208\,{\rm MeV/fm^3}.
\ee
Note, that for the isosymmetric matter, with equal numbers of $u$ and $d$
quarks, the resulting bag constant is 
$B_{0u}+B_{0d}\simeq 69\,{\rm MeV/fm^3}\simeq (152\,{\rm MeV})^{\,4}$\,.   

It is easy to verify that the following thermodynamic identities 
hold in the model
\bel{theri}
\rho=\partial_{\mu} P,\,\,\,\rhob=\partial_{\,\overline{\mu}} P\,.
\ee
In the mean--field approximation the entropy density
\begin{eqnarray}
s&=&\partial_{\,T} P=\partial_{\,T} P_K=\nonumber\\
&=&-\nu\,\sump\left[\,\ocn\ln{\ocn}+(1-\ocn)\ln{(1-\ocn)}
+ \ocn\to\ocnb\right]\label{entr}
\end{eqnarray}
is the same as for the mixture of ideal gases of quarks and 
antiquarks. Using the thermodynamic
relations for the free energy density 
\bel{fren}
{\cal F}\equiv\frac{F}{V}=\mu\,\rho+{\overline{\mu}}\,{\rhob}-P=e-Ts\,,
\ee
one can calculate the energy density of the \qq matter, $e$.
From~Eqs. (\ref{pres1}), (\ref{entr})--(\ref{fren}) one has
\bel{ende}
e\equiv\frac{E}{V}=e_K+\frac{\displaystyle G_V\rho_V^{\,2}}
{\displaystyle 2}+B(m)\,.
\ee
Here the kinetic part, $e_K$, again corresponds to the                         
mixture of quark and antiquark ideal gases:
\bel{kien}
e_K=\nu\sump E_{\bm{p}}\left(\,\ocn+\ocnb\right)\,.
\ee
The same expression for the energy density can be obtained
by averaging the energy-momentum tensor
\mbox{$<T^{00}>=<H>/V$}. According to~Eqs. (\ref{bagc}), 
(\ref{ende})--(\ref{kien}), the vacuum energy density 
is equal to zero: $e_{\rm vac}=B(m_{\rm vac})=0$\,.
   
From the above formulae it is easy to check that the following
differential relations 
\begin{eqnarray}
dP&=&\rho\,d\mu+\rhob\,d\overline{\mu} +s\,dT\,,\label{dpre}\\
d{\cal F}&=&\mu\,d\rho+\overline{\mu}\,d\rhob -s\,dT\,,\label{dfre}
\end{eqnarray}
hold for any process in a thermally equilibrated system.
Using~Eqs. (\ref{fren}), (\ref{dfre}) one can derive the relations
\begin{eqnarray}
P&=&\left(\,\rho\,\partial_{\displaystyle\rho}+
\rhob\,\partial_{\,\displaystyle\rhob}-1\,\right){\cal F}\nonumber\\
&=&\left(\,\rho_+\partial_{\displaystyle\rho_+}-1\right) {\cal F}+
\rho_V\partial_{\displaystyle\rho_V}{\cal F}\,,
\label{pres3}
\end{eqnarray}
where $\rho_+\equiv(\rho+\rhob\,)/2$. From this expression it is evident
that the pressure vanishes when the free energy per particle, 
\mbox{$F/\mbox{$<N+\overline{N}>$}={\cal F}/(\rho+\rhob\,)$}, has a minimum 
in the $\rho-\rhob$ plane. As shall be seen below, at low 
enough temperatures, a whole line of zero pressure states exists 
in this plane. But only one point on this line, namely the one corresponding
to the minimum of the free energy per particle should 
be regarded as a (meta)stable state. At $T=0$ 
this state corresponds to the minimum of the energy per particle,
\bel{eppa}
\epsilon\equiv\frac{\displaystyle E}{\mbox{$\displaystyle <N+\overline{N}>$}}=
\frac{\displaystyle e}{\rho+\rhob}\,.
\ee

Now we discuss the stability of the \qq matter. 
Thermodynamically stable matter may exist provided the
$2\times 2$ matrix 
$||\partial_{\displaystyle\rho_i} \partial_{\displaystyle\rho_j} {\cal F}||$
(with $\rho_1=\rho,\,\,\rho_2=\rhob$) is positive~\cite{Mul95}. 
If this condition is not fulfilled, the system will be unstable
due to a spontaneous growth of isothermal density fluctuations
(spinodal instability). 
From~\re{dfre} it is evident that the above stability condition
holds when the inequalities
\begin{eqnarray}
\partial_{\rho}\mu&>&0\,,\label{stab1}\\
\partial_{\rho}\mu\,\partial_{\,\rhob}\overline{\mu}
&>&\partial_{\,\rhob}\mu\,\partial_{\rho}\overline{\mu} \label{stab2}
\end{eqnarray}
hold simultaneously.
As will be shown below, outside the spinodal region, there exists
a wider region of metastable states where the coexistence
of two phases ($i=1,\,2$) with different densities and masses
is thermodynamically favourable. 
This is possible when the Gibbs conditions 
\begin{eqnarray}
\mu_1 &=& \mu_2\,,\label{mug}\\
\overline{\mu}_1 &=& \overline{\mu}_2\,,\label{mubg}\\
P_1 &=& P_2,\,\label{preg}\\
T_1 &=& T_2\label{temg}
\end{eqnarray}
are satisfied. 

In the mixed phase, the condition of the baryon charge conservation can be
written as 
\bel{bcc}
\lambda \left(\rho_1-\rhob_1\,\right)+
(1-\lambda) \left(\rho_2-\rhob_2\,\right)=3\,N_B/V\,,
\ee
where $\lambda\leq 1$ is the volume fraction 
of the phase 1 and $N_B$ is the total baryon
number of the \qq system. From Eqs.~(\ref{mug})--(\ref{bcc})
one can see that for $N_B\neq 0$ the pressure of the mixed state
will in general be $\lambda$--dependent, i.e. the Maxwellian
construction ($P(\lambda)={\rm const}$ at $T={\rm const}$) will be
violated. The situation is similar to the liquid--gas 
phase transition in nuclear matter with fixed baryonic and electric
charges~\cite{Gle92,Mul95}.
We would like to stress here that by imposing the condition of chemical 
equilibrium,~\re{cemc}, one normally puts the system outside the region
of the phase transition.

In the baryon--free \qq matter ($<N-\overline{N}>=0$) we 
have $\rho=\rhob$ and $\mu=\overline{\mu}$ (see~\re{norc}). In this case 
Eqs.~(\ref{stab1})--(\ref{bcc}) become much simpler.
In particular, the stability conditions,  
Eqs.~(\ref{stab1})--(\ref{stab2}), are equivalent   
to the requirement that the isothermal compressibility is 
positive:
\bel{itco}
\left(\partial_\rho P\right)_{\,T}=2 \rho\,(\partial_\rho\mu)_T>0\,.
\ee
Here we have used Eqs.~(\ref{dpre}), (\ref{stab1}).

It is interesting to note that one more instability 
may occur in a dense baryon--rich system due to the presence of strong
vector potential. At sufficiently high baryonic
density a finite \qq droplet becomes unstable with respect to  
the spontaneous creation of \qq pairs\footnote{
This phenomenon is analogous to the spontaneous electron--positron
pair production in strong electromagnetic fields~\cite{Sch51}.
}. 
This process becomes possible when
the vector field exceeds a certain critical value ~\cite{Mi93b}
\bel{vinc}
G_V|\rho-\rhob|>m+m_{\rm vac}\,.
\ee
When this condition is satisfied, the upper energy levels of
the Dirac sea reach the bottom of the positive energy continuum.
A rough estimate of minimal baryon density necessary to produce
a \qq pair can be obtained by omitting the first term in
the r.h.s. of~\re{vinc}. In the case $f=u, d$ this gives
the following estimate
for the critical baryon density $\rho^{\,c}_B$\,:
\bel{crid}
\rho^{\,c}_B=\frac{1}{3}(\rho-\rhob)\,\goo\,\frac{m_{\rm vac}}
{3\,G_V}\simeq\,6.3\rho_0\,,
\ee
where $\rho_0=0.17\,{\rm fm}^{-3}$ is the equilibrium density of 
normal nuclear matter.
In the above estimate we have used our standard choice of
the model parameters.
At larger $G_V$ the critical densities will be lower. 

Before going to the results, let us consider the special case 
of zero temperature. Taking the limit $T\to 0$ in~\re{ocnq}
and using~\re{norc} we have
\bel{ocnl}
n_{\bm{p}}\to\theta\,(\mu_R-E_{\bm{p}})
=\theta\,(p_F-p)\theta\,(\mu_R-m)\,,
\ee
where
\bel{fmom}
p_F=\sqrt{\mu_R^2-m^2}=\left(\frac{6\pi^2\rho}{\nu}\right)^{1/3}
\ee
is the Fermi momentum of quarks. From~\re{fmom} and the analogous expression
for antiquarks one gets the explicit formulae for the chemical potentials
at $T=0$\,:
\begin{eqnarray}
\mu&=&\sqrt{m^2+p_F^2}+G_V\,(\rho-\rhob)\,,\label{cpzt1}\\  
\overline{\mu}&=&\sqrt{m^2+\overline{p}_F^2}-G_V\,(\rho-\rhob)\,.\label{cpzt2}
\end{eqnarray}
From Eqs.~(\ref{gap1}), (\ref{scad1}) and (\ref{ocnl}) we arrive at
the following gap equation at zero temperature:
\vspace*{1mm}
\bel{gapl}
\frac{m_0-m}{G_S}=\frac{\nu}{8\pi^2}
\left[\,p_F^{\,3}\,\Psi\,'\left(\frac{m}{\,p_F}\right)
+\overline{p}_F^{\,3}\,\Psi\,'\left(\frac{m}{\,\overline{p}_F}\right)
-\Lambda^3\Psi\,'\left(\frac{m}{\Lambda}\right)\right]\,.
\vspace*{1mm}
\ee     

Two comments are in order here. First, from~Eqs.~(\ref{cpzt1})--(\ref{cpzt2})
one can see that the condition of chemical equilibrium, (\ref{cemc}),
can not be satisfied at $T=0$\,. At zero temperature only pure
baryon ($\rhob=0$) or antibaryon ($\rho=0$) matter should be considered as 
chemically equilibrated. Second, according to~\re{gapl}, 
the quark mass $m$ becomes smaller than
$m_0$ at sufficiently high $p_F$ or $\overline{p}_F$. 
However, the applicability  
of the NJL model is questionable at very high densities. 
Indeed, as discussed in Ref.~\cite{Kle92}, the cutoff momentum 
$\Lambda$ has to exceed all characteristic particle momenta.
In particular, the inequality 
$\Lambda>{\rm max}\,(p_F, \overline{p}_F)$ must not be violated.

The explicit formulae for the
energy density and the pressure at $T=0$ are given by 
the relations:
\begin{eqnarray}
e&=&\mu\,\rho+{\overline{\mu}}\,{\rhob}-P\nonumber\\
&=&\frac{\nu}{8\pi^2}
\left[\,p_F^{\,4}\,\Psi\left(\frac{m}{\,p_F}\right)+
\overline{p}_F^{\,4}\,\Psi\left(\frac{m}
{\,\overline{p}_F}\right)\right]
+\frac{G_V\rho_V^2}{2}+B(m)\,.\label{end2}
\end{eqnarray}
The gap equation,~\re{gapl}, corresponds to the minimum of $e$
at fixed $\rho$ and $\rhob$\,.

\section{Results}

In this section we present the results of numerical calculations.
They were performed by using the parameter set
from~\re{modp}. One should bear in mind that in some cases
the results are rather sensitive to the particular choice of model parameters.
These cases will be discussed in more detail. Special attention
will be paid to qualitative effects of chemical non--equilibrium
which, as we shall see later, strongly influence the thermodynamic
properties of \qq matter. 

First, we study the non--strange
matter consisting of light ($f=u,d$) quarks and antiquarks. 
The effects of isotopic asymmetry are disregarded i.e. we assume that 
$\rho_u=\rho_d$ and $\rhob_u=\rhob_d$. Second, we consider 
also the $s\overline{s}$ matter  with arbitrary densities of
quarks and antiquarks. In accordance with the SU(3) flavour 
symmetry of the interaction Lagrangian, the same 
coupling constants are used for strange and
non--strange systems.

Let us discuss first the properties of the non--strange \qq matter
at zero temperature. As noted above, chemical equilibrium
at $T=0$ can be realized only at $\rhob=0$ (in systems with 
positive baryon charge)\,. In particular, the   
baryon--free matter, with equal densities of quarks and antiquarks,
can not be chemically--equilibrated at $T=0$\,.   
By using the formulae of Sec.~3 we calculated
the quark mass $m$, chemical potentials $\mu,\,\overline{\mu}$, 
the pressure $P$
and energy per particle $\epsilon$ as functions of densities $\rho$            
and $\rhob$\,. The results are given in 
Figs.~1--7\,\footnote
{Unless stated otherwise, for systems with $f=u,d$ we denote 
by $P$ the total pressure of $u$ and $d$ quarks:
$P\equiv P_u+P_d=2P_u$\,. In the same case $\rho$ ($\rhob$) denotes  
the total density of light quarks (antiquarks), i.e. 
$\rho=\rho_u+\rho_d=2\rho_u$\,. When necessary,  
the single--flavour values are introduced by the subscript $f$\,.
}.     

Figure~1 shows the density dependence of $\epsilon, \mu$ and $m$
for baryon--free \qq matter composed of light quarks. The rapid drop
of the quark mass reflects the restoration of chiral
symmetry at high densities. The calculation shows that
$\epsilon (\rho)$ has a maximum at a relatively 
small quark density $\rho=\rho_B\ll\rho_{\,0}$. At a higher density, 
$\rho=\rho_A$, the energy per particle reaches a minimum
$\epsilon=\epsilon_A < \epsilon (0)=m_{\rm vac}$. Therefore,
the metastable state of baryon--free \qq matter 
is predicted at zero temperature. This state has
zero pressure (see below Fig.~3) and its existence is possible 
only in a chemically non--equilibrium system.

In the considered case the vector
density $\rho_{\,V}=0$ and therefore the density 
dependence of $\epsilon=e_f/2\rho_f$   
is determined by the balance between
the first kinetic term in~\re{end2} and the bag part $B(m)$\,.
At $\rho\to 0$ one has $m\simeq m_{\rm vac}$ and 
$B(m)\simeq\frac{1}{2}B\,''(m_{\rm vac})\,(m-m_{\rm vac})^2$\,. 
The explicit expression for $B\,''(m)$ follows 
from~Eqs.~(\ref{bagc}), (\ref{phim1}). The condition 
of vacuum stability (see the footnote to~\re{gapv}), 
$B\,''(m_{\rm vac})>0$, can be easily verified. Substituting 
\mbox{$p_F=\overline{p}_F=(\pi^2\rho_f)^{1/3}$} into~\re{end2}, in
the limit $p_F\ll m$ one has
\bel{eldl}
\epsilon\simeq m+\frac{3}{10}\,\frac{p_F^2}{m}+\frac{B(m)}{2\rho_f}\,.
\ee
Minimizing this expression with respect to $m$, one finds that
at low $\rho_f$ the quark mass decreases linearly with density:
\bel{mldl}
m\simeq m_{\rm vac} -\frac{2\rho_f}{B\,''(m_{\rm vac})}\,.
\ee
From Eqs.~(\ref{eldl})--(\ref{mldl}) one can get the approximate 
formulae
\bel{eld1}
\epsilon\simeq m_{\rm vac}+
\frac{3}{10}\,\frac{p_F^2}{\,m_{\rm vac}}-\frac{\rho_f}
{B\,''(m_{\rm vac})}\,.
\ee   

One can see that at low densities Eqs.~(\ref{mldl})--(\ref{eld1}) 
qualitatively reproduce the results shown in Fig.~1. Indeed, 
the second term on the r.h.s. of~\re{eld1} is proportional 
to~$\rho^{2/3}$. At small~$\rho$ this term dominates and $\epsilon$ 
increases with density reaching a maximum at the point~B.
From Eqs.~(\ref{fren}), (\ref{pres3}) at $\rho=\rhob,\,T=0$\,,
one can obtain the relations 
\bel{thrl}
P/2\rho=\rho\,\partial_\rho\epsilon=\mu-\epsilon\,,
\ee       
which show that  $P=0$ and 
$\mu=\epsilon$ at the extrema of $\epsilon$\,.

At $\rho>\rho^{(B)}$, the energy and pressure start to decrease due to
the chiral symmetry restoration effects. At sufficiently high densities,
the quark mass approaches the bare mass. In lowest order
one can take $m\simeq m_0\ll p_F$ and substitute $m=m_0$
into the expressions for $P$ and $e$ given by~\re{end2}.
This gives the following approximate formulae
\bel{hdes}
\mu\simeq p_F,\,\,\epsilon\simeq\frac{3}{4}\,p_F+\frac{B_0}{2\rho_f}\,,
\ee
where $B_0\equiv B(m_0)$\,is the bag constant. 
From Eqs.~(\ref{thrl})--(\ref{hdes}) one can see
that the pressure of a symmetric \qq system changes its sign 
at the point A, where the kinetic (Fermi--motion) term and the bag
part balance each other\footnote
{According to these estimates, the origin of the above
energy minimum is similar to that suggested in the MIT bag model 
to explain hadron properties in vacuum~\cite{Kap81}.
}:
\bel{pes1}
P_f^{(A)}\simeq\frac{1}{2}\rho_f\,p_F-B_0=0\,.
\ee
At much higher densities the bag contribution may be neglected
and we obtain the limit of a massless ideal Fermi--gas:  
$P\simeq\frac{1}{3}\,e\simeq\frac{1}{2}\rho\,p_F\propto\rho^{4/3}$\,.       

Substituting  $B_0=B_{0u}$ from~\re{bagc1} we get the estimate
$\rho_A=2\rho_u^{(A)}\simeq 3.0\,\rho_0$\,. This  
is close to the density of the metastable state
obtained by the numerical calculation, which gives the values
\bel{lmst}
\rho_A\simeq 2.93\,\rho_0,\,\,\epsilon_A=\mu_A\simeq 0.268\,{\rm GeV}\,.
\ee
As will be seen below, 
the value $\epsilon_A$ is the absolute minimum of
$\epsilon\,(\rho,\,\rhob)$ in the whole $\rho-\rhob$ plane.
It is interesting to note that the minimum energy 
value $\epsilon_A$ is below the
vacuum mass,~$m_{\rm vac}$. However, this state is
in fact metastable due to the possibility of the annihilation,
$\qq\to n\,\pi\,, n=2,3,\ldots$\,.   

The ''chemically--equilibrated'' case of baryon--rich matter 
with $\rhob=0$ is considered in Fig.~2. A comparison 
with the preceding figure
shows that the quark mass drops with the quark density slower
than in the baryon--free matter. Although the $\rho$--dependence
of $\epsilon$ is rather flat, for the present parameter set 
the energy per particle has no extrema 
in the considered case.  Formally one can use the same procedure
to estimate the behaviour of $\epsilon$ at $\rho\to 0$ 
as was suggested above for the baryon--free matter.
The resulting expression may be obtained  
by replacing the last term in Eq.~(\ref{eld1}) 
by $\,[\,G_V-1/B\,''(m_{\rm vac})\,]\,\rho_f/2$. Using this 
expression one can find a fictitious maximum of $\epsilon$
at $\rho\sim 6\rho_0$ which is not found in the numerical
calculation. At such a high density, however,
the assumptions $m\simeq m_{\rm vac}$, $p_F\ll m$
used in the above low density expansion are not valid. Therefore,
in the baryon--rich case the model  
predicts a qualitatively different equation 
of state than that for symmetric \qq matter.

Figure~3 shows the density dependence of the pressure
in the $u, d$ matter at $T=0$. Again, in the baryon--rich
case, $P(\rho)$ is monotonically increasing function. On the other hand, 
the behaviour of the pressure in the symmetric system is non--trivial:
there is a region of negative pressures at $\rho_B<\rho<\rho_A$.
The interval with the negative compressibility ($\partial_\rho P<0$)
exists between points B and C. One can see from Eqs.~(\ref{pres3}), 
(\ref{stab1}), that the system is mechanically unstable at
$\rho_B<\rho<\rho_C$. The appearance of regions with negative $P$
and $\partial_\rho P$ are necessary conditions for the existence of 
bound states and for the first order phase transitions,
respectively.       

One should bear in mind that the presence or 
absence of states with negative $P$ and $\partial_\rho P$ 
in the quark matter with $\rhob = 0$ is quite sensitive to the 
choice of the model parameters. The similar 
conclusion was made in Refs.~\cite{Kli90a,Bub96}. 
By taking small values of the 
vector coupling constant $G_V$ and/or quark bare mass $m_0$, one 
can find the bound states and phase 
transitions even in baryon--rich systems. This is clear from 
Fig.~4 which shows the sensitivity of the equation of state
to the choice of $G_V$ and $m_0$ in $u, d$ matter with 
$\rhob = 0$\,. In the case $m_0=0$, 
bound states of the baryon--rich matter are possible only if 
$G_V\loo\,0.35\,G_S$. On the other hand, in absence of the vector interaction,
i.e. at $G_V=0$, the model predicts bound states
even for bare masses \mbox{$m_0\sim 5-8\,{\rm MeV}$,} corresponding to
correct values of the pion decay constant $f_\pi$\,.
However, at the realistic choice of the model parameters
we do not find any phase transitions or bound states in the
chemically equilibrated matter.
The stable quark droplets have been predicted in 
Ref.~\cite{Rag98} within the generalized NJL model.
As discussed in Ref.~\cite{Bub96}, the prediction is questionable
since it was implicitly assumed that $G_V=0,\,m_0=0$.     
With the same caution one should regard the results 
of Refs.~\cite{Asa89,Cug96,Kli97}, where the possibility of phase transitions 
in baryon--rich \qq matter was studied within the NJL model,
either without vector interaction or by assuming unrealistically small
values of~$G_V$\,.        

Figures 5(a), (b) show the model predictions for arbitrary 
\qq systems with light quarks at $T=0$.
The contours of equal energy per particle and pressure 
in the $\rho-\rhob$ plane are given in
Fig.~5(a) and 5(b), respectively. Only the vertical and horizontal 
axes of this plane 
correspond to chemically equilibrated states at zero temperature.  
One can see that symmetric systems with 
$\rho=\rhob\simeq 3\,\rho_0$ indeed have the non--trivial 
minimum of the energy per particle
and zero pressure. Shaded areas indicate the spinodal regions
where the conditions (\ref{stab1})--(\ref{stab2}) are violated.
From Fig.~5(b) one can see that 
at $T=0$  practically all unstable states are located inside 
the contour $P=0$\,. The presence of a spinodal instability shows to the
possibility of the phase coexistence in a wider density region.
We shall discuss this issue in more detail at the end 
of this section.  

The results for $s\overline{s}$ systems are presented in Figs.~6, 7.
According to Fig.~6, the density dependence of $m,\,\epsilon$ and
$\mu$ in the symmetric $s\overline{s}$ matter is qualitatively 
the same as for non--strange systems. The model predicts 
a metastable state (point A) with the parameters
\bel{smst}
\rho=\rhob\simeq 6.26\,\rho_0,\,\,\epsilon\simeq 0.451\,{\rm GeV}\,.  
\ee              
Although the assumption $m\ll p_F$ is not so well--justified
as for light quarks,~\re{pes1} gives the estimate 
$\rho_A\simeq 5.8\,\rho_0$ which is close to the numerical result.
The noticeably higher density of the metastable state in the $s\overline{s}$
case is due to the higher bag constant of strange quarks. Indeed,
from~\re{bagc1} one can see that $B_{0s}/B_{0u}\sim 6$\,.
\re{pes1} leads to the estimate 
\mbox{$0.5\,(B_{0s}/B_{0u})^{3/4}\simeq 1.9$} for the ratio
of densities of the metastable states in strange and non--strange
systems, rather close to the value obtained from
numerical calculations.  

As can be seen from Fig.~7(a), the symmetric $s\overline{s}$
system with the parameters given by~\re{smst}, has 
the minimum energy as compared to other points in the 
$\rho-\rhob$ plane. According to Fig.~7(b), the model does not 
predict states negative $P$ and $\partial_\rho P$
in baryon--rich matter with $\rhob=0$. A~special study
with different vector coupling constants shows that, 
in the latter case, the states  with $P<0$ exist at $G_V/G_S<0.175$. 
The analogous calculation shows that negative values of 
$\partial_\rho P$ are possible if $G_V/G_S<0.215$. 
The shaded area again shows the region of spinodal instability.
For strange \qq matter the instability region extends
to higher densities as compared to the non--strange matter.    
    
Let us now discuss nonzero temperatures. 
\re{norc} yields the relations 
\begin{eqnarray}
\frac{\rho}{T^3}&=&I_0\left(\frac{m}{T},
\frac{\mu_R}{T}\right)\,,\label{norc1}\\
\frac{\rhob}{T^3}&=&I_0\left(\frac{m}{T},
\frac{\overline{\mu}_R}{T}\right)\,,\label{norc2}
\end{eqnarray}
connecting the  quark ($\mu_R$) and antiquark ($\overline{\mu}_R$) 
reduced chemical potentials with the densities 
$\rho$ and $\rhob$\,. The dimensionless function
\bel{densi}
I_0(x,y)=\frac{\nu}{2\pi^2}\int\limits_0^\infty {\rm d}t
\frac{t^2}{\exp{(\sqrt{x^2+t^2}-y)}+1}
\ee
appears from the integration of the particle occupation
numbers in momentum space.

Using Eqs.~(\ref{pres1}), (\ref{gap2}), one may write
the gap equation in the form
\bel{gap3}
(\partial_m P_K)_{T,\,\mu_R,\,\overline{\mu}_R}=B\,'(m)\,.
\ee     
Substituting the expression for $P_K$ from~\re{presk}, we have 
\bel{gap4}
-\frac{B\,'(m)}{m\,T^2}=I_1\left(\frac{m}{T},\frac{\mu_R}{T}\right)+
I_1\left(\frac{m}{T},\frac{\,\overline{\mu}_R}{T}\right)\,.
\ee
Here $I_1(x,y)$ differs from $I_0(x,y)$ by the replacement 
${\rm d}t \to {\rm d}t/\sqrt{t^2+x^2}$ in the r.h.s. of~\re{densi}.
The above equation is the analogue of~\re{gapl} for the case 
$T\neq 0$\,.

According to Eqs.~(\ref{cemc}), (\ref{rce1})--(\ref{rce2}), 
chemical equilibrium in \qq matter is achieved if 
\mbox{$\mu_R=-\overline{\mu}_R$}\,\footnote{
Therefore, the symmetric matter, where $\mu_R=\overline{\mu}_R$
(see  Eqs.~(\ref{norc1})--(\ref{norc2})), may be 
chemically equilibrated only if $\mu_R=\overline{\mu}_R=0$, 
which in this case is equivalent to $\mu=\overline{\mu}=0$.
}. 
Using this condition and Eqs.~(\ref{norc1})--(\ref{norc2}),
(\ref{gap4}) it is easy to see that the densities
$\rho$ and $\rhob$ are not independent 
in chemically equilibrated systems: the density 
of antiquarks at a given T is fixed
by the density of quarks, and vice versa. Figs. 8(a), (b)
show the dependence $\rhob\,(\rho)$ at various temperatures
for the non--strange and strange matter in chemical equilibrium.
We see that up to rather high temperatures, $T\sim 100$\,MeV
for $f=u,d$ and $T\sim 150$\,MeV for $f=s$, the equilibrium 
values of $\rhob$ or $\rho$ are very small. This is especially
evident for the regions far from the diagonal $\rhob=\rho$.
 
Figures~9 and 10 represent the mass and pressure isotherms in symmetric 
\qq matter with light and strange quarks. These results
are obtained by the simultaneously solving 
Eqs.~(\ref{norc1})--(\ref{norc2}), (\ref{gap4}) with respect to
$m,\,\mu_R$ and $\overline{\mu}_R$. 
Figs.~9(a), (b) show that the quark mass increases
with $T$ at a fixed $\rho$. Such a behaviour is a specific feature
of chemically non--equilibrated systems. It can be qualitatively understood,
at least in the limit $m\ll T$. Indeed, from the definition
of $B(m)$ it is easy to show that the l.h.s of~\re{gap4} decreases
with $m$\,. On the other hand, by using~\re{norc1}  
one can see that the r.h.s. of the same equation increases with $\rho/T^3$\,.
This follows from the fact that $I_i (0,y),\,i=1,2$ are increasing
functions of $y$\,. This explains the above mentioned behaviour of $m\,(T)$\,.
Figs.~9--10 show also the density dependence of $m$ and 
$P$ in the limit of chemical equilibrium, i.e. at $\mu=\overline{\mu}=0$. 
According to Eqs.~(\ref{norc1}), (\ref{gap4}), 
(\ref{pres1})--(\ref{presk}), in this limit $m,\,\rho$ and $P$ are 
fully determined by the temperature only.  

As can be seen from Figs.~10(a), (b), at a fixed density the 
pressure increases with $T$. Similar to the zero temperature case,
one can find the spinodal instability regions
with negative isothermal compressibilities $(\partial_\rho P)_T$. 
The calculations show that the stability condition (\ref{itco}) 
in symmetric \qq matter
is violated at temperatures $T<T_c$ where
\bel{crte}
T_c\simeq\left\{\begin{array}{lll}76\,&{\rm MeV},\,&f=u,\,d,\\
92\,&{\rm MeV},\,&f=s.\end{array}\right.
\ee
At lower temperatures the first order phase transition 
is possible. This means that the formation 
of the mixed phase is thermodynamically favourable.

The coexisting phases are characterized (for symmetric \qq matter)
by different densities $\rho_1<\rho_2$
and quark masses $m_1>m_2$, but they have equal chemical potentials
($\mu_1=\mu_2\equiv\mu$) and pressures. Formally
such a situation is possible in the regions 
of temperature and chemical potential 
where the gap equation has several solutions for the quark mass.
Using the formulae of Sec.~3, one arrives at the following
set of equations for $m_1,\,m_2$ and $\mu$ at fixed~$T$:
\vspace*{2mm}
\begin{equation}
\left\{\begin{array}{ll}
P_K(m_1,\mu,T)-B(m_1)=P_K(m_2,\mu,T)-B(m_2)\,,\label{bin1}\\[1mm]
\partial_{m_1}P_K(m_1,\mu,T)=B\,' (m_1)\,,\label{bin2}\\[1mm]
\partial_{m_2}P_K(m_2,\mu,T)=B\,' (m_2)\,.\label{bin3}
\end{array}\right.\\[2mm]
\ee
The values of $\rho_i\,(i=1,2)$ are obtained by substituting
$m=m_i$ and $\mu_R=\mu$ into~\re{norc1}. The
density points  $\rho_1$ and $\rho_2$
are represented by dotted lines in Figs.~10(a),~(b).
These lines constrain the two--phase ``binodal'' regions shaded
at the same plots. At each temperature the states with a negative 
compressibility are situated inside the  
interval $\rho_1<\rho<\rho_2$\,. Similar to the well--known
liquid--gas phase transition, the states betwen the spinodal
and binodal lines are metastable with respect to the 
heterophase decomposition of \qq matter. 

As can be seen from Fig.~10, states with zero pressure 
exist at sufficiently low temperatures. 
The maximal temperature when the existence of such states is still possible 
equals \mbox{$T_m\simeq 49\,(67)$\,MeV} for 
non--strange (strange) symmetric matter. At $T<T_m$ the states with
$P=0$ are in the metastable density region. Thin lines
in Figs.~10(a), (b) represent the density dependence of the pressure
in chemically equilibrated baryon--free matter. The points where this 
line intersects the pressure isotherms give 
the corresponding values of $P$ and $\rho$ for the case of
chemical equilibrium. These points are located outside of the 
binodal region.       

More detailed information concerning the phase transition
in the baryon--free matter is given in Fig.~11.
The phase coexistence lines in the $\mu-T$ plane are shown
both for non--strange and strange baryon--free matter. 
One can clearly see that states 
with $\mu=\overline{\mu}=0$ do not belong to the phase transition lines.
Therefore, the discussed phase trasition has essentially
non-equilibrium nature. It does not occur in chemically 
equilibrium systems.

Figures~12(a), (b) represent the contour lines of $\epsilon$ and $P$
for the non--strange \qq mixture at \mbox{$T=50$\,MeV.}  
This temperature is close to the maximum temperature for the existence
of zero pressure states. The comparison with Fig.~5 shows that 
the spinodal region occupies a smaller part of the $\rho-\rhob$ plane
than in the case $T=0$\,. From Fig.~8(a) we conclude
that at such a low temperature the equilibrium quark--antiquark 
densitiies lie outside (below) the spinodal boundary.
An analysis of the binodal boundaries in the $\rho-\rhob$ plane will be 
given in a future publication.    

\section{Summary and discussion}

We have demonstrated that
the NJL model predicts the existence of metastable states of \qq matter
out of chemical equilibrium. These novel states are most 
strongly bound for symmetric (net baryon--free) systems 
with equal numbers of quarks and antiquarks. For realistic model parameters
there exists a region of quark chemical potentials and temperatures
where a first order chiral phase transition is possible. 
It is shown that for the same set of parameters the first
order phase transition does
not occur in a chemically equilibrated system. This
conclusion may also be valid for other models of strongly interacting
matter. Up to now both the chiral and the deconfinement
transition in hadronic and quark--gluon matter have been studied under
the assumption that chemical equilibrium is fulfilled. 
As discussed above, this is most likely not the case in 
relativistic heavy--ion collisions, at least at some stages 
of matter evolution.

Let us briefly discuss possible signatures of the phase transitions
and of the bound states predicted in this paper.
Measurements of excitation functions of particle spectra  
in heavy--ion collisions have been suggested~\cite{Ste98}  
to detect a possible critical point in the \qq matter phase diagram. 
In principle, the same idea can be applied to the search 
for these non--equilibrium phase transitions.
However, an easier task might be to look experimentally for
the decay of metastable \qq droplets in nuclear
collisions. It is clear that the formation of  
such droplets would be possible if 
dense and relatively cold baryon--free matter is created
at an intermediate stage of the reaction. Such conditions
may be realized at high bombarding energies,
presumably, at RHIC and LHC. Selecting 
events with unusually high pion multiplicity and low 
$<p_T>$ may increase the probability to observe such multiquark droplets. 
Considerations similar to those suggested 
in Ref.~\cite{Mis95} show that the subsequent hadronization
of droplets may give rise to narrow bumps in hadron 
rapidity distributions. These bumps may be observable in an 
event--by--event analysis of $\pi,\,K$ and $\phi$ spectra.
It is also obvious, that the yields
of $B\overline{B}$ pairs ($B=N,\,\Lambda,\,\Sigma,\,\Xi,\,\Omega\,\ldots$)
will be enhanced if they are produced through the decay 
of multiquark--antiquark systems.  

The understanding of the possible decay channels of the predicted
metastable states is of great importance. In non--strange 
systems, the strongest decay channel should be the annihilation 
of \qq pairs into pions.
This channel is energetically open at least at the surface
of the \qq droplet. Indeed, according to~\re{lmst}, the average energy
released in the annihilation of a \qq pair is  
$\mu+\overline{\mu}=2\,\epsilon_A\simeq 3.8\,m_\pi$. 
This is sufficient to create up to three pions.
The largest rate of annihilation is apparently given by the $\qq\to 2\pi$
channel. Indeed, the one--pion annihilation is suppressed 
in homogeneous matter due to the energy--momentum conservation.
On the other hand, the relative probability of the $\qq\to 3\pi$ channel
should be small due to the limited phase space volume available. 
However, the pion properties
in the non--equilibrium \qq medium have not yet been investigated. 
It may happen that pion--like
excitations disappear due to the Mott transition~\cite{Kle92}.
It is also possible that the in--medium pion mass is increased as compared 
to the vacuum value. In both cases the annihilation inside the \qq 
droplet will be suppressed. 

The situation in a metastable 
$s\overline{s}$ droplet may be even more complicated.
It is interesting that the direct annihilation of the
$s\overline{s}$ pairs into hadrons is suppressed.  
Indeed, according to ~\re{smst}, the energy available in 
annihilation of a $s\overline{s}$ pair 
is equal on average to \mbox{$\mu+\overline{\mu}=2\,\epsilon\simeq 0.9$\,\,GeV.} 
This is lower than the threshold energies of the final
hadronic states: $\phi\,(1020\,{\rm MeV}),\, 
K\overline{K}\,(990\,{\rm MeV})$\,and $\rho\pi\,(910\,{\rm MeV})$\,. 
In this case, annihilation
is possible only for energetic $s\overline{s}$ pairs from the tails of
the thermal distribution.  
The only open channel is $s\overline{s}\to 3\,\pi$\,. 
From the analogy to the $\phi\to 3\,\pi$\,decay, which has 
the partial width $\Gamma_{\phi\to 3\pi} < 80$\,keV, 
one can conclude that this channel
should be quite weak.

However, in the deconfined \qq matter, the strong decay channels 
$s\overline{s}\to u\,\overline{u},\,d\,\overline{d}$ are possible,
which may reduce significantly the life times of $s\overline{s}$ droplets.
These processes will lead to the conversion of $s\overline{s}$ matter
into a system where all three flavours are present. If sufficiently
long times were available, chemical equilibrium between the 
flavours would be established. It will be interesting to check whether an
admixture of \qq pairs in strange baryon--rich systems
is energetically favourable as compared with ordinary 
stranglets~\cite{Far84,Gre87}. Detailed calculations of the formation 
probability and life times of metastable \qq droplets are possible 
only within a dynamical approach. 

One should investigate also the role of the mesonic 
degrees of freedom which have been disregarded in the
present work. It is well known that mesons appear in the NJL model 
as bound states of a quark and an antiquark. The influence of
mesonic degrees of freedom in chemically equilibrated \qq 
matter has been studied in Ref.~\cite{Zhu94}.
It was shown by using RPA that the contribution of mesons
may be important at low quark densities.

In the present
calculation the flavour mixing six--fermion interaction was neglected.
This interaction has been introduced earlier~\cite{Kli90b}
to reproduce the mass splitting of $\eta$ and $\eta'$ mesons. 
It has been shown in Ref.~\cite{Bub96} that the flavour--mixing interaction 
only slightly changes the EOS of non--strange \qq matter at $T=0$.
It will be interesting to investigate the role of this interaction 
in the case when chemical equilibrium is violated.

\section*{Acknowledgments}

We thank M.I. Gorenstein and S. Schramm for useful discussions.
One of us (I.N.M.) is grateful to the Alexander von Humboldt Foundation 
for the Research award. Two of us (I.N.M. and L.M.S.) 
thank the Niels Bohr Institute, University of Copenhagen, 
and the Institute for Theoretical Physics, University 
of Frankfurt am Main, for the kind hospitality and financial support.
This work has been  supported by 
the Graduiertenkolleg Experimentelle and Theoretische 
Schwerionenphysik, GSI, BMBF and DFG.

\section *{Figure captions}
\newcommand{\Fig}[2]{\noindent{FIG.~#1.~}
\parbox[t]{15cm}{\baselineskip 24pt #2}\\[5mm]}
\Fig{1}
{Energy per particle, chemical potential and constituent quark mass 
in symmetric matter with equal numbers of light quarks 
and antiquarks at zero temperature.
$\rho$ is the total density of $u$ and $d$ quarks. 
Point A shows the position of the metastable state.
}  
\Fig{2}
{The same as in Fig.~1, but for asymmetric $u,\,d$ matter 
with zero density of antiquarks.
}
\Fig{3}
{Pressure of \qq matter with light quarks at zero temperature.
The solid (dashed) line corresponds to the baryon--free (rich) system.  
}
\Fig{4}
{Regions in the $m_0-G_V$ plane where states with negative pressure 
(dark shading) and  negative compressibility (light shading)
exist in baryon--rich ($\rhob=0$) quark matter at $T=0$\,.
Cross corresponds to parameters used in the present model. 
}
\Fig{5}
{Contours of equal energy
per particle (in GeV, upper part) and equal pressure 
(in GeV/fm$^3$, lower part) 
in \qq matter with light quarks at zero temperature.
$\rho\,(\rhob)$ is total density of $u$ and $d$ quarks (antiquarks).
Shading indicates the regions of spinodal instability.
}
\Fig{6}
{The same as in Fig.~1, but for symmetric $s\overline{s}$ matter.
}
\Fig{7}
{The same as in Fig.~5, but for $s\overline{s}$ matter.
}
\Fig{8}
{Quark and antiquark densities, $\rho$ and $\rhob$,  
in chemichally equilibrated \qq matter \mbox{($\overline{\mu}=-\mu$)} 
at different temperatures (shown in MeV). Upper and lower part
correspond to the non--strange and strange matter,
respectively.
}
\Fig{9}
{Mass isotherms for non--strange (upper part) and strange
(lower part) symmetric \qq systems. 
Temperatures 
are shown in MeV near the corresponding curves. 
Thin solid lines show the density dependence of quark mass in 
chemically equilibrated matter ($\mu=\overline{\mu}=0$)\,.
}
\Fig{10}
{Pressure isotherms for baryon--free \qq matter. 
Upper (lower) part represents model predictions 
for non--strange (strange) systems. Thin solid lines 
show the pressure of chemically equilibrated matter.
The binodal regions are shaded.
}
\Fig{11}
{Phase coexistence lines in the $\mu-T$ plane 
for symmetric non--strange (solid line) and strange 
(dashed line) \qq matter.
}
\Fig{12}
{The same as in Fig.~5, but for the temperature $T=50$ MeV.
}
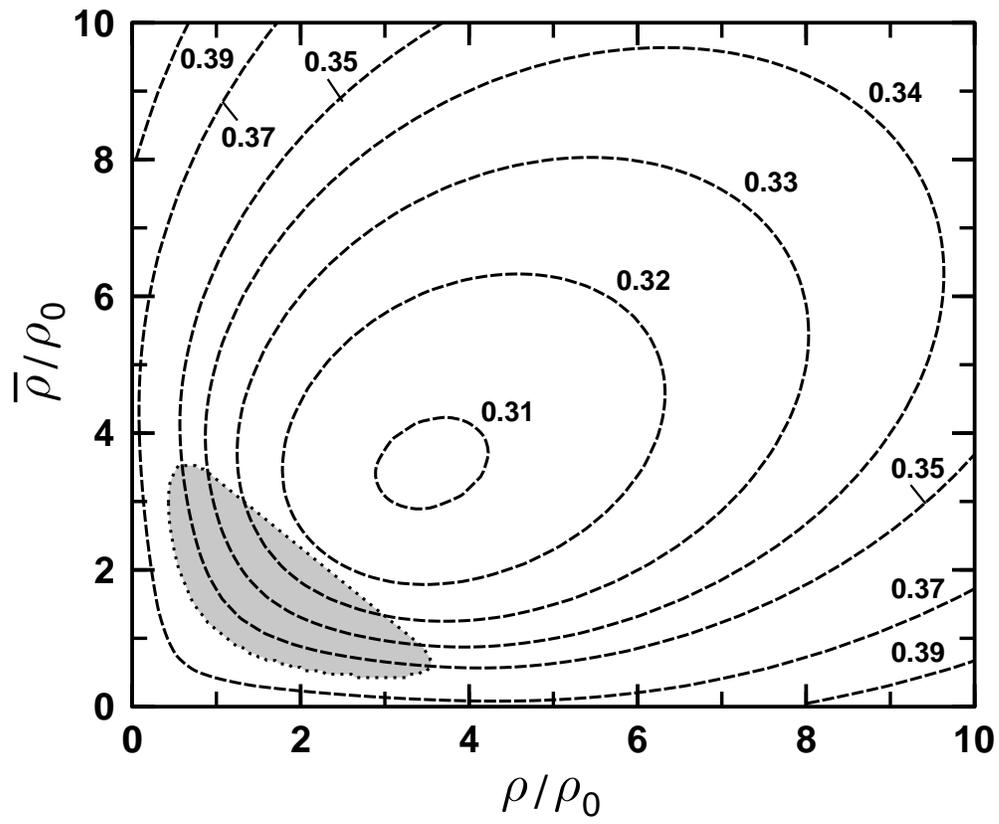
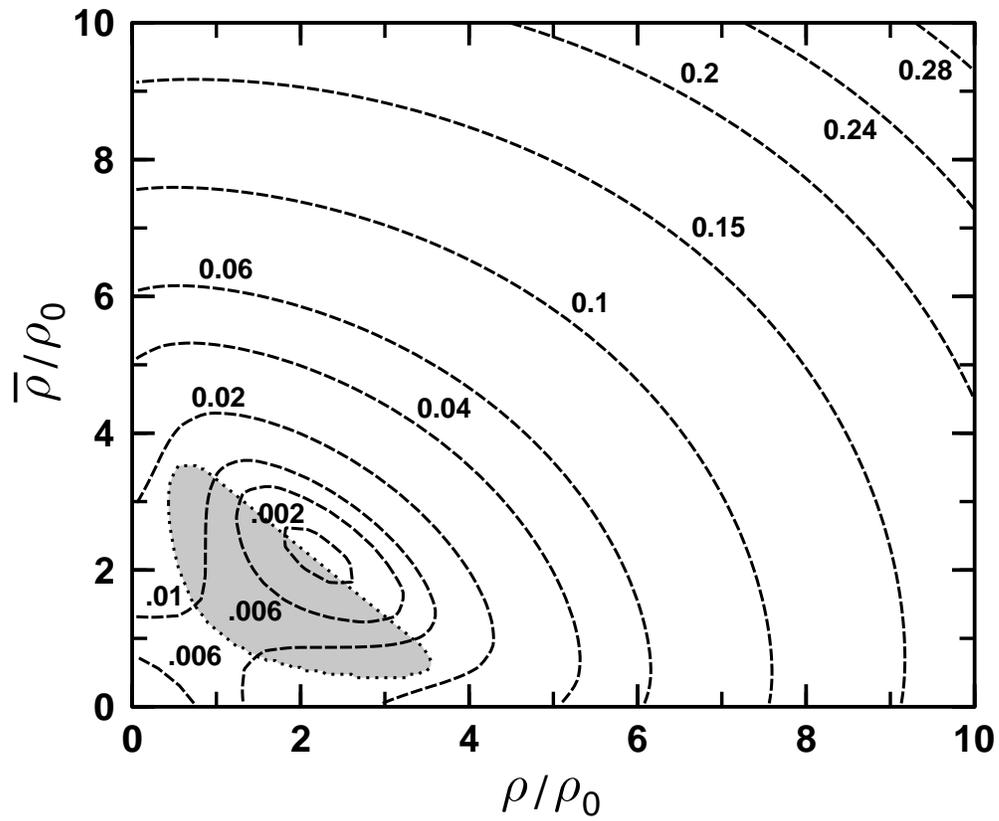
\end{document}